\documentclass[aps,prd,twocolumn,groupedaddress,showpacs]{revtex4-1}

\usepackage[T1]{fontenc}

\usepackage{amsmath}
\usepackage{mathrsfs}
\usepackage{amssymb}
\usepackage{graphicx}

\usepackage{enumitem}
\usepackage{tikz}
\usepackage{circuitikz}
\usepackage{scalerel}
\usepackage[compat=1.0.0]{tikz-feynman}
\usepackage{epsfig}

\usepackage{hyperref}
\hypersetup{
    colorlinks=true,
    linkcolor=blue,
    filecolor=magenta,      
    urlcolor=blue,
    citecolor=blue
}

\begin{document}

\title{Simulation of the process $e^+e^- \rightarrow W^+W^-$ with the heavy right-handed neutrino exchange at 1 TeV future lepton colliders}

\author{A. Drutskoy and E. Vasenin}

\affiliation{P.N. Lebedev Physical Institute of the Russian Academy of Sciences, Moscow 119991, Russia}

\date{\today}

\begin{abstract}

We study potential contribution of the heavy right-handed neutrino exchange in the process $e^{+}e^{-} \rightarrow W^{+}W^{-}$.
This process is sensitive to heavy neutrinos with masses larger than $\sqrt{s}$.
The Monte Carlo simulation of the studied process is performed assuming the seesaw type-I model,
where heavy right-handed neutrinos (heavy neutral leptons,  HNLs) are introduced in the leptonic sector. 
Within the Standard Model (SM), the process has a large cross section described by diagrams with 
$s$-channel $Z / \gamma$ exchange and $t$-channel active neutrino exchange. Respectively, the $t$-channel right-handed neutrino
exchange amplitude will interfere with these SM amplitudes.
However, the angular distributions of the $W$ boson production and decay are different for the right-handed neutrino and SM amplitudes. That can be used to evaluate potential HNL contribution using the extended likelihood method.
The simulation of the $e^{+}e^{-} \rightarrow W^{+}W^{-}$ process is performed 
at the 1 TeV center-of-mass energy and polarization ${\cal{P}}_{e^+e^-}$ of (20\%, $-80\%$),  which is a standard option for the future linear $e^+e^-$ International Linear Collider.
Both $W$ bosons are reconstructed from two hadronic jets.  Simulation of the SM background
processes is also done. The beam-induced backgrounds and the initial state radiation (ISR) effects are taken into account. 
The majority of background processes are effectively suppressed by the cuts
on the invariant masses of two and four jets. 
Finally, we obtain upper limits on the mixing parameter $|V_{eN}|^2$ as a function of $M(N)$.

\end{abstract}


\maketitle

\section{INTRODUCTION}

We study the process $e^+e^- \rightarrow W^+W^-$ assuming contribution of the heavy right-handed neutrinos (heavy neutral leptons, HNLs, $N$) by $t$-channel exchange.  
Heavy right-handed neutrinos appear in many beyond the
Standard Model models to explain small masses of the active neutrinos~\cite{PhysRevLett.81.1562,PhysRevLett.89.011301,PhysRevLett.90.021802}.
HNLs can generate Dirac masses of the active neutrinos with the Higgs mechanism~\cite{PhysRevD.96.095004,KANEMURA201166,Jana_2020}. However, this mechanism requires Yukawa coupling to be of $y_D \sim 10^{-12}$, which 
is difficult to explain. There are several studies within a generalized bottom-up effective field theory, in which right-handed neutrinos naturally appear~\cite{PhysRevD.94.055022,PhysRevD.96.015012,Li_2021,DELAGUILA2009399, physrev_80, physlett_95, physrev_76}. 
The most popular extensions of the lepton sector are within various seesaw models, which are widely discussed today~\cite{cai2018leptonnumberviolationseesaw}.

The process $l^-l^- \to W^-W^-$ ($l = e,\mu$) has been discussed in several 
papers~\cite{Asaka_2015,jiang2023searchingmajorananeutrinossamesign, samesign}, assuming opportunity for future same-sign electron or muon colliders. 
This process with $t$-channel HNL exchange is possible, if heavy neutrinos have a Majorana nature. In this process
the lepton number violates by two units, therefore it has no backgrounds in the Standard Model. 
Using our experimental procedures we can well reproduce the results obtained in these papers. However, the same-sign lepton colliders have a less
rich physics program that the opposite-sign versions~\cite{emem}. Therefore we study the similar process for the case of the
opposite-sign beams.

In contrast to processes with direct HNL production~\cite{samesign,M_ka_a_2022,Antonov_2023}, the $e^+e^- \rightarrow W^+ W^-$ process is sensitive to contributions of HNLs
with very large masses, $M(N) \gg \sqrt{s}$.
Within the framework of the seesaw type-I model, HNL masses can be large; therefore, it is important
to have experimental accessibility to the large HNL mass region. Since heavy right-handed neutrinos have large mass,
interference of the SM and HNL diagrams results in specific angular distributions
in the $W$ boson production and decays. We use these 
distributions within the extended likelihood method to evaluate the HNL contribution
to the process and to obtain upper limits on the HNL mixing parameters. The circular $e^+e^-$ colliders FCC-ee~\cite{FCC_tdr} and CEPC~\cite{CEPC_tdr} with collision energies up to 380 GeV are widely discussed today.
However, at the region $\sqrt{s} \sim 200\text{--}500~\text{GeV}$ the SM 
$e^+e^- \to W^+W^-$ cross section is $\sim 1-2~\text{pb}$ and the $t$-channel
HNL exchange contributions are too small for experimental searches. In the paper \cite{PhysRevD.57.2771} virtual HNL loop
contributions were calculated and found to be also very small in this $\sqrt{s}$ region.

Current experimental upper limits on the mixing parameters $|V_{lN}|$, $l = e, \mu, \tau$ have been discussed in \cite{Abdullahi_2023,sterile,hnlwhitemass}.
There are strict upper limits $|V_{eN}|^2 \lesssim  10^{-5}$ for the HNL masses less than the
$Z$ boson mass.  However in the region $M(N) > 100~\text{GeV}$, the upper limits, mostly obtained by the LHC, are weak, $|V_{eN}|^2 \gtrsim 10^{-2}$ for 
$M_N < 1~\text{TeV}$ and $|V_{eN}|^2 \gtrsim  1$ for $M_N > 1~\text{TeV}$. The latest results of searches for HNLs
performed by the LHC Collaborations are presented in~\cite{cmshnl1,cmshnl2,cmshnl3,cmshnl4,atlascollaboration2024searchheavyrighthandedmajorana}.
As we report in this paper, future lepton colliders can significantly improve the upper limits in the large HNL mass region up to ${\cal{O}}(10^4 - 10^5)~\text{GeV}$.

We study the process within the seesaw type-I model framework.
Right-handed neutrinos are naturally introduced 
in the seesaw type-I model. They transform
as a singlet under the SM gauge group and interact with the SM leptons through a Yukawa coupling.
The Lagrangian of this interaction is given by
\vspace{-0.2cm}
\begin{equation}
    {\cal{L}}_N = - \overline{L} Y^D_{\nu}\tilde{H}N_R - \dfrac{1}{2}\overline{(N^c)}_LM_RN_R + \text{H.c.},
    \label{seesaw_lagr}
\end{equation}
where $L$ and $H$ are the left-handed lepton and Higgs doublets, respectively. Once $H$ settles on the vacuum expectation value $\left\langle H \right\rangle = v_0/\sqrt{2}$, neutrinos acquire Dirac masses
$m_D = Y^D_{\nu}v_0/\sqrt{2}$, and Eq. (\ref{seesaw_lagr}) transforms into
\vspace{-0.2cm}
\begin{equation}
    \begin{split}
    {\cal{L}}_N \ni -\dfrac{1}{2}(\overline{\nu}_Lm_DN_R+\overline{(N^c)}_Lm_D^T(\nu^c)_R + \\
    + \overline{(N^c)}_LM_RN_R) +\text{H.c.}
    \end{split}
\end{equation}

To obtain the mass eigenstates one introduces a unitary transformation
\vspace{-0.2cm}
\begin{equation}
    \begin{pmatrix}
        \nu \\
        N^c 
    \end{pmatrix}_L = U \begin{pmatrix}
        \nu_m \\
        N^c_{M}
    \end{pmatrix}_L,
\end{equation}
where $m$ is the active neutrino mass, and $M$ is the HNL mass. After unitary transformation,
we obtain the neutrino mass matrix 
\begin{equation}
    U^{\dagger}\begin{pmatrix}
        0 & m_D \\
        m_D^T & M_R
    \end{pmatrix} U = \begin{pmatrix}
        m_{\nu} & 0 \\
        0 & M_N
    \end{pmatrix}.
\end{equation}
In the limit of large $M_R$ it results in the ratio between the masses of active and heavy neutrinos
\begin{equation}
    m_{\nu} = -m_D M^{-1}_R m_D^T, \ \ \ \ M_N \approx  M_R.
\end{equation} 

From the latest equation, it is seen that large HNL masses favor small masses of active neutrinos; therefore, it is
important to test the large HNL mass region experimentally.

This study includes simulation and reconstruction of the process $e^+e^- \rightarrow W^+W^-$ to produce respective Monte Carlo (MC) data samples. 
The following technical analysis of these samples is performed to estimate upper
limits on the mixing parameters $|V_{eN}|^2$ as a function of the HNL masses. The generator \footnotesize WHIZARD \normalsize version 3.1.5~\cite{Kilian_2011,moretti2001omegaoptimizingmatrixelement}
is used to calculate matrix elements and to simulate phase space. The HeavyN model~\cite{Alva_2015,PhysRevD.94.053002,Pascoli_2019,Atre_2009} is additionally
included in the generator to calculate the amplitude with the HNL exchange. Pythia6~\cite{pythia} is used for hadronization and $W$ boson
decay simulation and hadronization. This setup is the one used in the International Linear Collider (ILC) technical design report~\cite{ilc_tdr}. The \footnotesize DELPHES~\normalsize\cite{2014} package is used for fast detector simulation and event reconstruction. Finally, the analysis of reconstructed
events is performed by the \footnotesize ROOT \normalsize package~\cite{root}.

\section{Experimental procedures}

\subsection{Monte Carlo simulation and event reconstruction}

The signal process  $e^+e^- \rightarrow W^+W^-$ is described by the diagrams with the $s$-channel $Z/\gamma$ exchange and by the diagrams with the $t$-channel neutrino (light or heavy)
exchange as shown in Figs.~1(a) and 1(b).
For simplicity, we assumed that only one heavy neutrino is present.

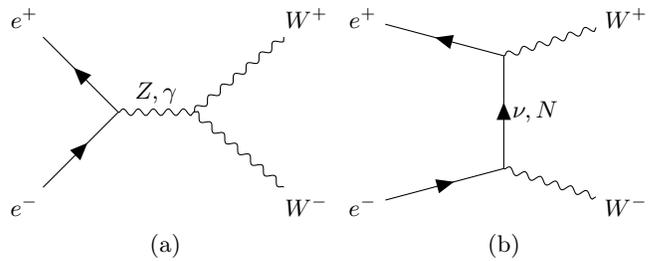
\begin{figure}[h]
    \centering
    \begin{tikzpicture}[]
        \begin{feynman}
            \vertex (E1) {$e^+$};
            \vertex [below = 2.5 cm of E1] (e1) {$e^-$};

            \vertex [right = 3.75cm of E1] (W1) {$W^+$};
            \vertex [below = 2.5 cm of W1] (w1) {$W^-$};

            \vertex [right = 1.25 cm of e1] (a);
            \vertex [above = 1.25 cm of a] (b);
            \vertex [right = 1 cm of b] (c);

            \vertex [right = 4.5 cm of E1] (E12) {$e^+$};
            \vertex [below = 2.5 cm of E12] (e12) {$e^-$};

            \vertex [right = 3.5cm of E12] (W12) {$W^+$};
            \vertex [below = 2.5 cm of W12] (w12) {$W^-$};

            \vertex [right = 1.875 cm of e12] (a2);
            \vertex [above = 0.5 cm of a2] (b2);
            \vertex [above = 1.5 cm of b2] (c2);

            \vertex[right = 1.875 cm of e1] (zxc);
            \vertex[right = 1.875 cm of e12] (zxc1);

            \vertex[below = 0.25cm of zxc] (capture1) {(a)};
            \vertex[below = 0.25cm of zxc1] (capture2) {(b)};
            
            \diagram* {
                (e1) -- [fermion] (b) -- [fermion] (E1),
                (b) -- [photon, edge label = ${Z, \gamma}$] (c),
                (w1) -- [boson] (c) -- [boson] (W1),

                (e12) -- [fermion] (b2) -- [fermion, edge label' = ${\nu, N}$] (c2) -- [fermion] (E12),
                (c2) -- [boson] (W12),
                (b2) -- [boson] (w12),
            };
        \end{feynman}
    \end{tikzpicture}

    \caption{\label{feyn_t}The $e^+e^- \rightarrow W^+W^-$ diagrams with $s$-channel (a) and with $t$-channel (b) exchanges}
    \label{diagrams}
\end{figure}

At the first step, the \footnotesize WHIZARD \normalsize generator is used to calculate the matrix elements
and to perform phase space Monte Carlo simulation. The matrix elements
are calculated for every diagram describing the process. The amplitudes
are then summed up taking into account the interference
between the diagrams.
The event generation is performed at the center-of-mass energy $\sqrt{s} = 1~\text{TeV}$
and the polarization ${\cal{P}}_{e^+e^-}$ of (20\%, $-80\%$),
which is a standard option for the future
ILC \cite{adolphsen2013internationallinearcollidertechnical}. 
The initial state radiation effects are taken into account using \footnotesize WHIZARD \normalsize in-built package.
Beam-induced backgrounds simulated using the CIRCE2 package are added to the $e^+e^- \rightarrow W^+W^-$ events.
The subsequent decays of the $W$ bosons into hadronic jets and following
hadronization is performed by Pythia6.

For a process described by several diagrams, \footnotesize WHIZARD \normalsize allows one to calculate a single matrix element for a particular diagram or a set of diagrams. We calculate the cross sections corresponding to the HNL diagram only to see the behavior of the cross section as a function of the HNL mass.
The cross sections are calculated for $|V_{eN}|^2 = 0.0021$ and different beam polarization options, as shown in Fig. 2.
The contribution of the HNL diagram becomes smaller at the large $M(N)$ region, respectively the upper limits on the mixing parameters become weaker 
with the HNL mass increase. The largest cross section is obtained for the beam polarization ${\cal{P}}_{e^+e^-}$ of (20\%, $-80\%$).

\begin{figure}[h]
    \includegraphics[width = 0.5\textwidth]{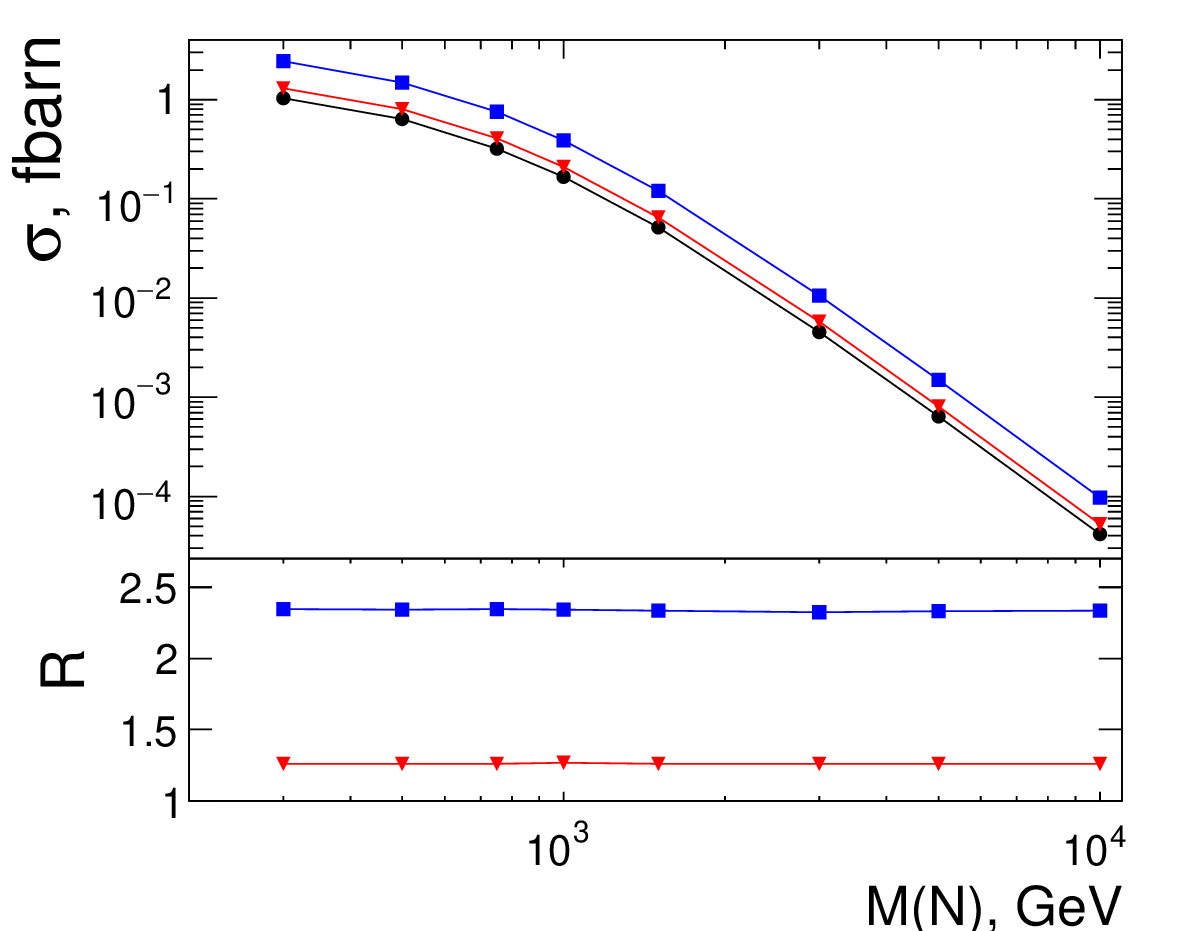}
    \caption{\label{cross}The cross sections of the $e^+e^-\rightarrow W^+W^-$ process with only HNL diagram taken into account for different
    beam polarizaton options and $\sqrt{s} = 1~\text{TeV}$ are shown. Blue squares correspond to ${\cal{P}}_{e^+e^-}$ of (20\%, $-80\%$), red triangles to ${\cal{P}}_{e^+e^-}$ of (20\%, 80\%) and black  circles to unpolarized beams. The lower part shows the ratio $R$ of the cross sections for the cases of the polarized and unpolarized beams.}
\end{figure}

Further detector simulation and event reconstruction are performed by the \footnotesize DELPHES \normalsize package. A preinstalled detector card is required in \footnotesize DELPHES \normalsize for the fast detector simulation.
We use the ``ILCgen'' card that performs simulation corresponding to the ILC detector presented in the technical design report \cite{behnke2013internationallinearcollidertechnical4}. This card describes responses of different subdetectors and provides realistic resolutions in the energy, momentum and other physical parameters.

We select events with four jets and no leptons in the final state.
To reconstruct jets in the hadronic calorimeter, we use the Valencia clustering algorithm \cite{Boronat_2018}
with the default parameters $R = 1.0$, $\beta = 1.0$ and $\gamma = 0.5$. If a $W$ boson has a large momentum,
the two produced hadronic jets overlap and, by default, are
considered as one ``fat jet'' by the jet clustering algorithm.
However, we force the algorithm to find exactly four jets, and therefore do not
lose events with fat jets that are forced to be split into two jets.
We test all jet pair combinations and choose the $W$ candidates
with the mass closest to the nominal mass of the $W$ boson. In case of fat jets,
the individual jet parameters can be somewhat incorrect,
but the combination of the two jets accurately
reproduces the $W$ boson candidate parameters.  

\subsection{Preselections}

The initial preselections require the transverse momenta
and pseudorapidities of the hadronic jets
to be within the detector acceptance,
\begin{equation}
    P_t(j) > 20~\text{GeV},\hspace{0.2cm}|\eta(j)| < 2.17.
\end{equation}

As the effects induced by the exchange of a right-handed neutrino are enhanced in the central region,
both $W$ bosons are required to be central
\begin{equation}
    |\cos \theta| < 0.7,
\end{equation} where $\theta$ is an angle between the direction of the $W$ boson and the beam.

These preselections are applied at the generator level
and significantly reduce the sizes of the studied event samples. 

\subsection{Backgrounds and cuts}

The studied backgrounds must have four jets and no leptons in the final state.
In addition, the background process
$e^+e^- \rightarrow q\bar{q}$ could produce two fat jets
reconstructed as four jets;
therefore, it also has to be considered.
The same procedure with the same parameters is used to simulate and reconstruct background events as it is done for the signal events.
The following background processes are generated, which are expected to be non-negligible in this study:

\begin{enumerate}[itemsep = -0.1cm]
    \item[(a)] $e^+e^- \rightarrow W^+W^-$ (with $s$-channel $Z / \gamma$ exchange and $t$-channel active neutrino exchange),
    \item[(b)] $e^+e^- \rightarrow W^+W^-\nu_e\bar{\nu_e}$,
    \item[(c)] $e^+e^- \rightarrow q\bar{q}$ (all flavors including $t\bar{t}$),
    \item[(d)] $e^+e^- \rightarrow ZZ$,
    \item[(e)] $e^+e^- \rightarrow HZ$.
\end{enumerate}

To distinguish between the signal and the background events, we use the following variables:
\begin{enumerate}[itemsep = -0.1cm]
    \item[(a)] $M(jj)$: the mass of a jet pair corresponding to the $W$ boson candidate;
    \item[(b)] $M(4j)$: the mass of four jets
    \item[(c)] $\sum p_x$ and $\sum p_y$: the sum of the $p_x$ or $p_y$ momentum components
      of all reconstructed final state objects in the event.
\end{enumerate}

To suppress most of the backgrounds except the process $e^+e^- \rightarrow W^+W^-$ the following cuts are applied:
\vspace{-0.1cm}
\begin{equation}
    70 < M(jj) < 90~\text{GeV},
\end{equation}
\vspace{-0.8cm}
\begin{equation}
    M(4j) > 600~\text{GeV},
\end{equation}
\vspace{-0.8cm}
\begin{equation}
    \left|\sum p_x\right| < 100~\text{GeV}, \hspace{0.5cm} \left|\sum p_y\right| < 100~\text{GeV}.
\end{equation}

The $M(jj)$ and $M(4j)$ distributions for the signal and background events are shown in Figs.~3 and 4, respectively.
The cuts on the $M(4j)$ and $\sum p_x / p_y$ variables are used to remove most
of the backgrounds with energetic unregistered particles in the final state.
The cut on $M(jj)$ is used to suppress backgrounds
where the two chosen hadronic jets do not correspond to a $W$ boson.
It is seen in Fig. 3 that this cut effectively suppresses most
of the background events. 
The numbers of events in the generated samples before and after applied cuts are shown in Table~I.
The HNL contribution is calculated for $|V_{eN}|^2 = 0.01$, $M(N) = 300~\text{GeV}$.
Also shown in Table~I is the SM contribution to $e^+e^- \to W^+W^-$.
The event yields are normalized to the integrated
luminosity of $1~\text{ab}^{-1}$. 

\begin{figure}[h]
    \includegraphics[width = 0.5\textwidth]{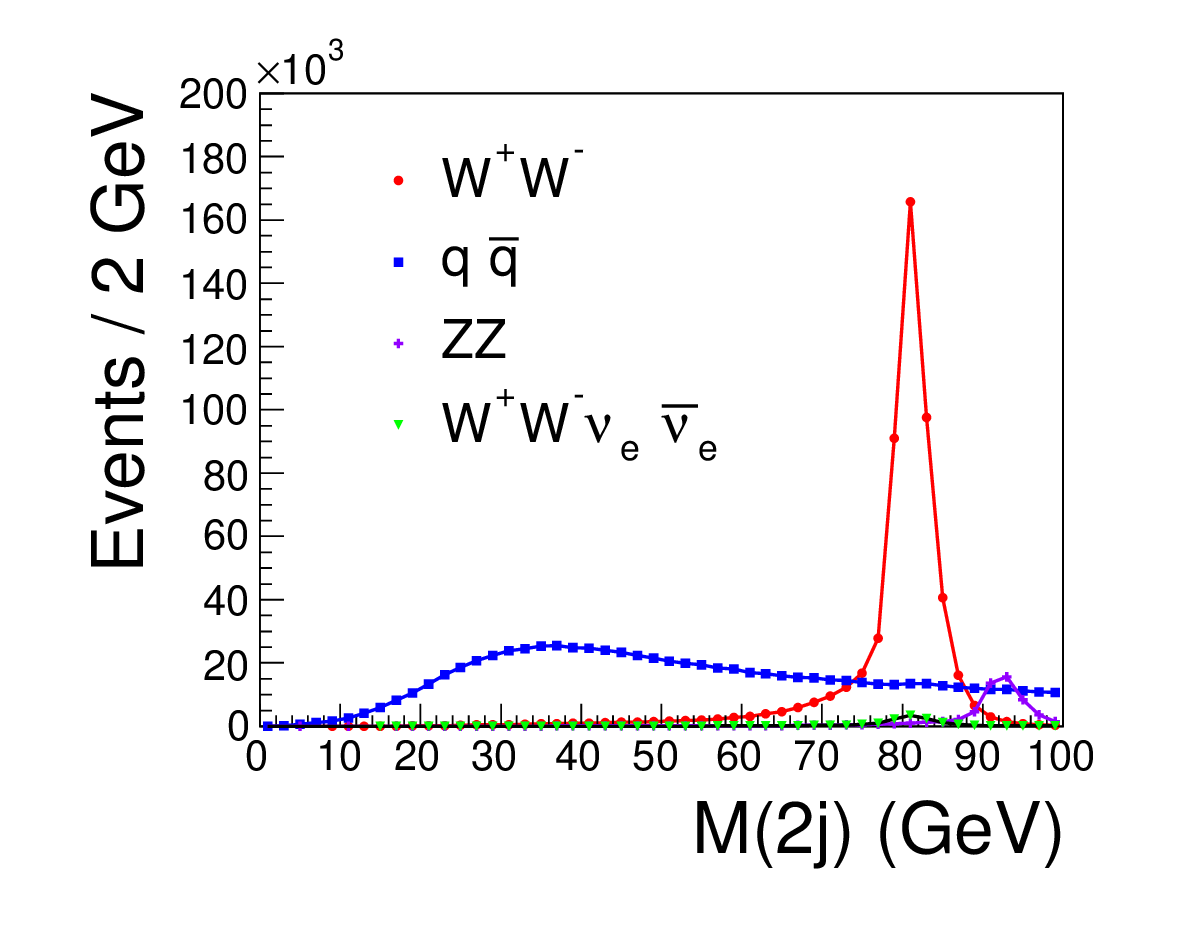}
    \caption{The $M(jj)$ distributions for the processes $e^+e^- \rightarrow W^+W^-$ (red circles), $e^+e^- \rightarrow q\bar{q}$ (blue squares),
    $e^+e^- \rightarrow ZZ$ (purple crosses), and $e^+e^- \rightarrow W^+W^-\nu_e\bar{\nu_e}$ (green triangles) for the integrated luminosity of $1~\text{ab}^{-1}$.}
\end{figure}

\begin{figure}[h]
    \includegraphics[width = 0.5\textwidth]{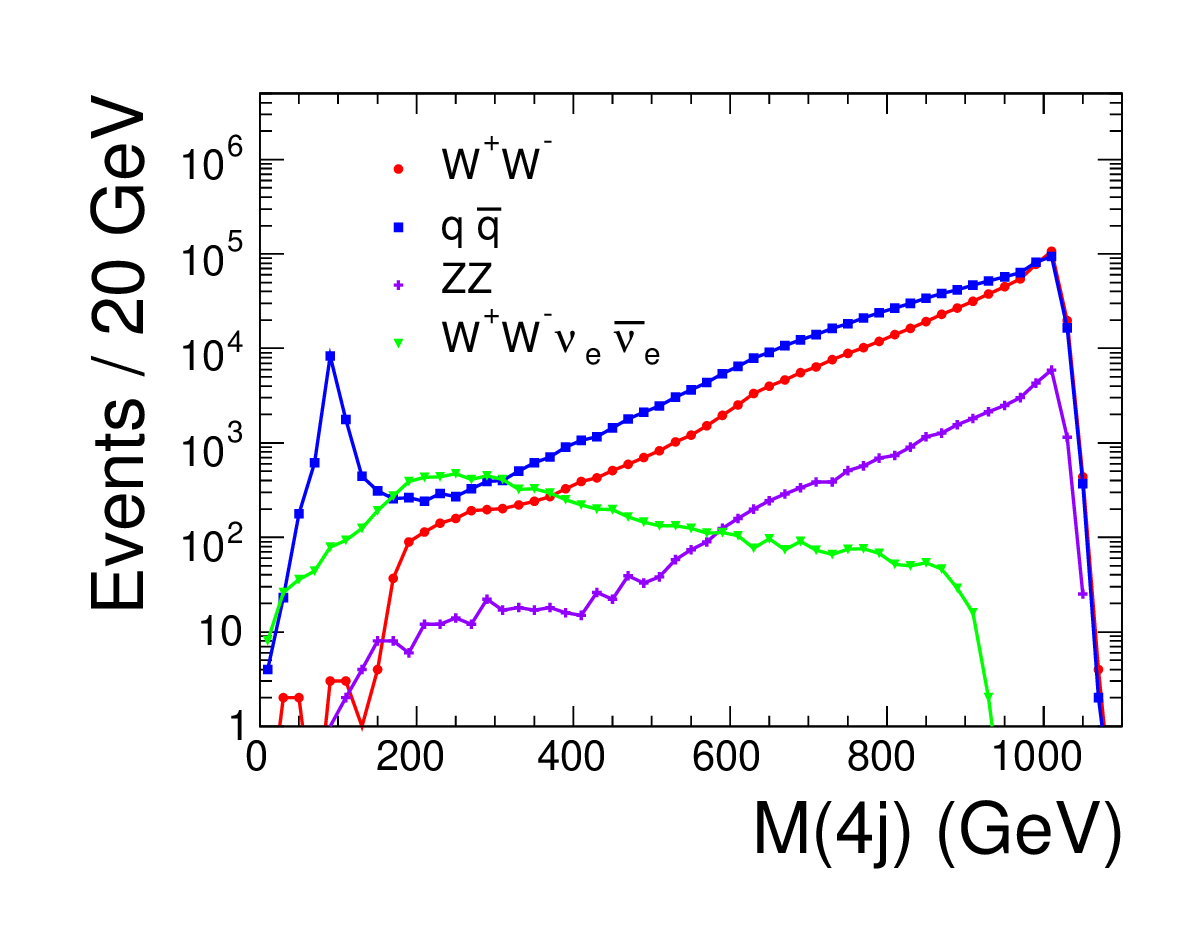}
    \caption{The $M(4j)$ distributions for the processes $e^+e^- \rightarrow W^+W^-$ (red circles), $e^+e^- \rightarrow q\bar{q}$ (blue squares),
    $e^+e^- \rightarrow ZZ$ (purple crosses), and $e^+e^- \rightarrow W^+W^-\nu_e\bar{\nu_e}$ (green triangles) for the integrated luminosity of $1~\text{ab}^{-1}$.}
\end{figure}

\begin{table}[h!]
    \caption{Numbers of events in the studied samples for $\mathcal{L}_{\rm int} = 1~\text{ab}^{-1}$,  $\sqrt{s} = 1~\text{TeV}$, ${\cal{P}}_{e^+e^-}$ = (20\%, $-80\%$), and preselections described above. We assume $|V_{eN}|^2 = 0.01$, $M(N) = 300~\text{GeV}$ for (SM+HNL) process.}
    \begin{center}
    \begin{ruledtabular}
        \begin{tabular}{l c c c}
            Process & Events after & Events after & $\varepsilon~(\%)$ \\
            $e^+e^- \rightarrow$ & preselections & all cuts & \\ \hline 
            $W^+W^-$ (SM+HNL) & 661674 & 531906 & $80.3$ \\ %\hline
            $W^+W^-$ (SM) & 525524 & 417889 & $79.5$ \\ %\hline
            $q\bar{q}$ & 716317 & 5005 & $0.7$ \\ %\hline
            $ZZ$ & 29419 & 1175 & $4.0$ \\ %\hline
            $W^+W^-\nu_e\bar{\nu_e}$ & 6041 & 328 & $5$\\ %\hline
            $HZ$ & 1037 & 6 & $0.5$ \\
        \end{tabular}
        \end{ruledtabular}
    \end{center}
    \end{table}

\subsection{Angular distributions}

The $e^+e^- \rightarrow W^+W^-$ process is described by five angles $\theta, \theta_1, \theta_2, \phi_1, \phi_2$
as shown in Fig.~5. Three of them, $\theta, \theta_1, \theta_2$,
are used to distinguish the SM+HNL contribution from the SM-only contribution.
The production angle $\theta$ is the angle between the $W$ boson direction
and the $e^-$ beam direction in the laboratory system.
The decay angles $\theta_1$ and $\theta_2$ are the angles between the jet direction
and the $e^-$ beam direction in the $W$ boson rest frame. $\theta_1$ is the decay angle of the $W$ boson with the larger angle with the $e^-$ beam direction and $\theta_2$ is the decay angle of the second $W$ boson.
Because of the large mass of the heavy neutrino, the distributions
of the variables $\theta, \theta_1, \theta_2$ for the SM and HNL contributions
are different.
Therefore, we can use these distributions to estimate the significance of the HNL contribution. 
The distributions of angles $\phi_1$ and $\phi_2$ between the production and decay planes
for the signal and background events are very similar
and are not used in this study. 

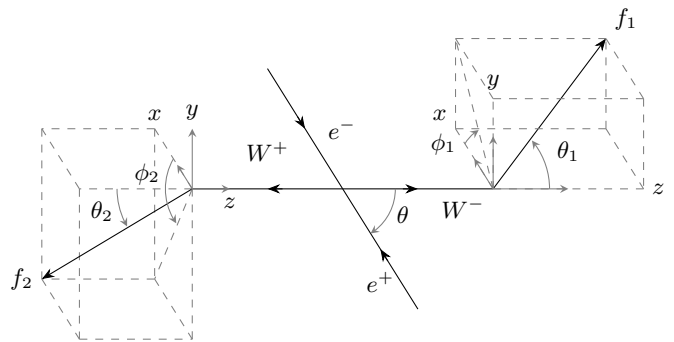
\begin{figure}[h]
    \begin{tikzpicture}
        \draw[-Stealth] (0,0) -- (1,0);
        \draw[] (0.8,0) -- (2,0);
        \draw[-Stealth] (2,0) -- (3.5,2) node[anchor = south west]{$f_1$};
        \draw[color = gray, -stealth] (2,0) -- (3,0);
        \draw[color = gray, dashed] (2,0) -- (4,0) node[anchor = west, color = black]{$z$};
        \draw[color = gray, -stealth] (2,0) -- (2, 0.7);
        \draw[color = gray, dashed] (2,0) -- (2, 1.2) node [anchor = south, color = black] {$y$};
        \draw[color = gray, dashed] (2,0) -- (1.5, 0.8) node [anchor = south east, color = black]{$x$};
        \draw[color = gray, dashed] (1.5, 0.8) -- (1.5, 2);
        \draw[color = gray, dashed] (1.5, 2) -- (2, 1.2);
        \draw[color = gray, dashed] (1.5, 2) -- (3.5, 2);
        \draw[color = gray, dashed] (2, 1.2) -- (4, 1.2);
        \draw[color = gray, dashed] (4, 1.2) -- (4,0);
        \draw[color = gray, dashed] (4, 1.2) -- (3.5, 2);
        \draw[color = gray, -stealth] (2, 0) -- (1.75, 0.4);
        \draw[color = gray, dashed] (1.5,0.8) -- (3.5, 0.8);
        \draw[color = gray, dashed] (3.5, 0.8) -- (3.5, 2);
        \draw[color = gray, dashed] (3.5, 0.8) -- (4, 0);
        \draw[color = gray, dashed] (2, 0) -- (1.5, 2);
        \draw[color = gray, -stealth] (1.625, 0.6) node[anchor = east, color = black]{$\phi_{1}$} arc (150:120:0.5);
        \draw[color = gray, -stealth] (2.75, 0) arc (0:40:1);
        \draw[] (1.6, 0) node[anchor = north]{$W^{-}$};
        \draw[] (3, 0.5) node{$\theta_1$};

        \draw[] (0.6, -0.96) -- (0,0);
        \draw[-Stealth] (1, -1.6) -- (0.5, -0.8) node[anchor = north, inner sep = 0.3cm]{$e^+$};
        \draw[color = gray, -stealth] (0.7,0) arc (0:-58:0.7);
        \draw[] (0.8,-0.35) node{$\theta$};

        \draw[] (-0.6, 0.96) -- (0,0);
        \draw[-Stealth] (-1, 1.6) -- (-0.5, 0.8) node[anchor = west, inner sep = 0.3cm]{$e^-$};

        \draw[-Stealth] (0,0) -- (-1,0);
        \draw[] (-0.8,0) -- (-2,0);

        \draw[color = gray, dashed] (-2, 0) -- (-3.5, 0);
        \draw[color = gray, dashed] (-3.5, 0) -- (-3.5, -2);
        \draw[color = gray, dashed] (-3.5, -2) -- (-2, -2);
        \draw[color = gray, dashed] (-2, -2) -- (-2, 0);
        
        \draw[color = gray, dashed] (-2, 0) -- (-2.5, 0.8) node [anchor = south, color = black]{$x$};
        \draw[color = gray, dashed] (-3.5, 0) -- (-4, 0.8);
        \draw[color = gray, dashed] (-2.5, 0.8) -- (-4, 0.8);
        \draw[color = gray, dashed] (-4, 0.8) -- (-4, -1.2);
        \draw[color = gray, dashed] (-2.5, 0.8) -- (-2.5, 0.3);
        \draw[color = gray, dashed] (-2.5, 0) -- (-2.5, -1.2);
        \draw[color = gray, dashed] (-2.5, -1.2) -- (-4, -1.2);
        \draw[color = gray, dashed] (-4,-1.2) -- (-3.5, -2);
        \draw[color = gray, dashed] (-2.5, -1.2) -- (-2, -2);

        \draw[] (-1, 0.5) node{$W^+$};

        \draw[-Stealth] (-2, 0) -- (-4, -1.2) node[anchor = east]{$f_2$};
        \draw[color = gray, -stealth] (-2,0) -- (-2.2, 0.32);
        \draw[color = gray, -stealth] (-2,0) -- (-2, 0.8) node[anchor = south, color = black]{$y$};
        \draw[color = gray, -stealth] (-2,0) -- (-1.5, 0) node[anchor = north, color = black]{$z$};

        \draw[color = gray, dashed] (-2,0) -- (-2.5, -1.2);
        \draw[color = gray, -stealth] (-3,0) arc(180:210:1);
        \draw[] (-3.2, -0.3) node{$\theta_2$};

        \draw[color = gray, -stealth] (-2.25, 0.4) arc (150:215:0.8);
        \draw[] (-2.6, 0.2) node{$\phi_2$};
        
    \end{tikzpicture}
    \caption{The angles used in the analysis of the process $e^+e^- \rightarrow W^+W^-$ with the subsequent $W$ boson decay into two 
    hadronic jets}
\end{figure}

The event simulation is performed for three different cases:  SM-only, HNL-only,
and SM+HNL.
The simulation of the $e^+e^- \rightarrow W^+W^-$ process with only the HNL diagram
included at the generator level is performed to demonstrate
the shapes of the respective angular distributions.
The distributions of $\cos \theta$ and  $\cos \theta_{1,2}$ are shown in Figs.~6 and 7, respectively.
The figures demonstrate the difference in the shapes of the angular distributions for the SM and HNL diagrams.
In the simulation the parameters $|V_{eN}|^2 = 0.0225$ and $M(N) = 500~\text{GeV}$ are used. These parameters provide larger signals than the parameters used in Table I to make signals more visible.

\begin{figure}[h]
    \includegraphics[width = 0.5\textwidth]{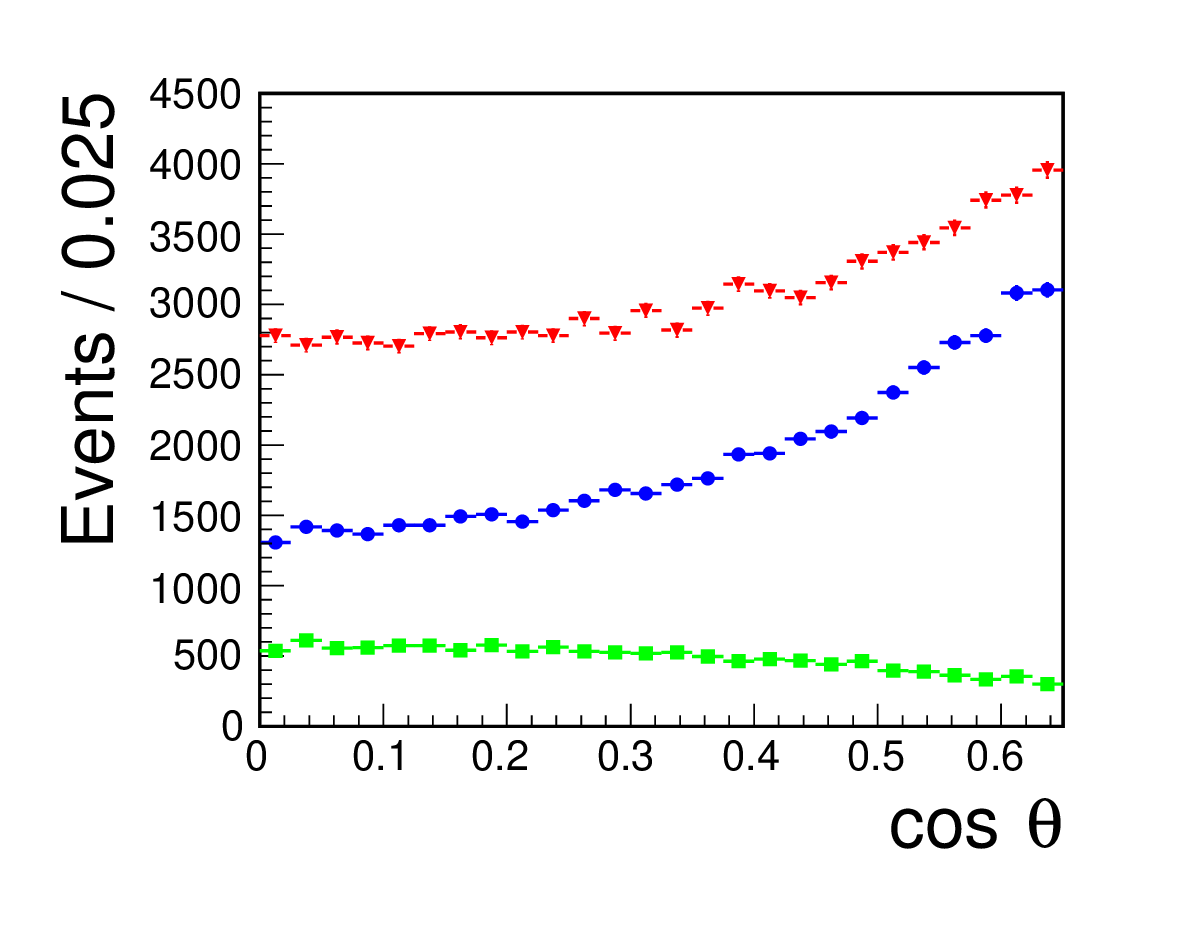}
    \caption{The $W$ production angular distributions for SM-only $e^+e^- \rightarrow W^+W^-$ (blue circles),
    HNL-only $e^+e^- \rightarrow W^+W^-$ (green squares),
    and the SM+HNL interference (red triangles) for the integrated luminosity of $1~\text{ab}^{-1}$.}
\end{figure}

\begin{figure}[h]
    \includegraphics[width = 0.5\textwidth]{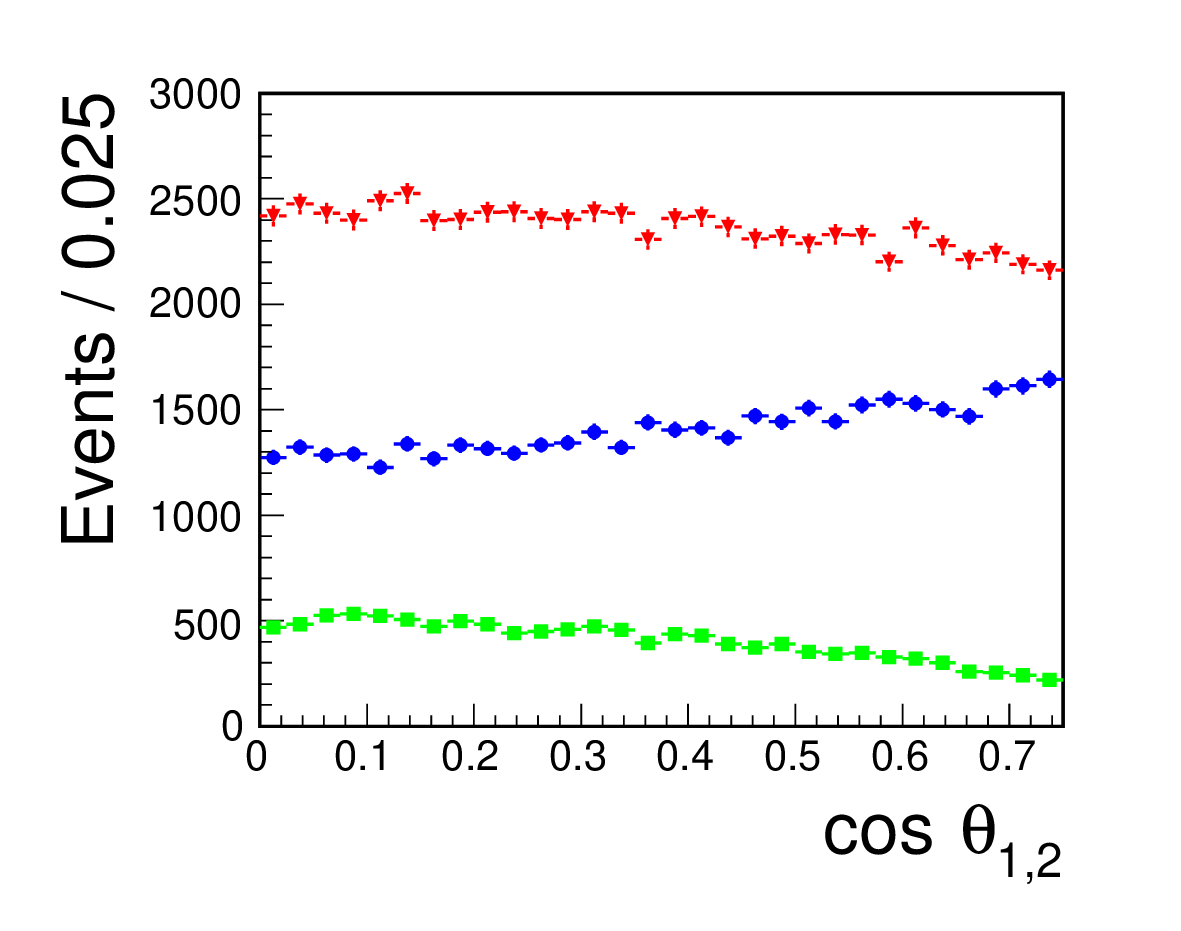}
    \caption{The $W$ decay angular distributions for SM-only $e^+e^- \rightarrow W^+W^-$ (blue circles),
      HNL-only $e^+e^- \rightarrow W^+W^-$ (green squares),
      and the SM+HNL interference (red triangles) for the integrated luminosity of $1~\text{ab}^{-1}$.}
\end{figure}

\section{Analysis}

To determine the upper limits on the mixing parameter we use the extended likelihood method.
The null hypothesis corresponds to the Standard Model case
without the HNL contribution.
The non-null hypothesis includes a nonzero HNL amplitude
interfering with the SM amplitudes. 
The significance of the HNL contribution is calculated as 
\vspace{-0.2cm}
\begin{equation}
    \label{significance}
    \textit{S} = -2\ln \dfrac{{\cal{L}}_0}{{\cal{L}}_1},
\end{equation}
where ${\cal{L}}_0$ is the likelihood function for the null hypothesis,
and ${\cal{L}}_1$ is the likelihood function for the non-null hypothesis.
The values of the likelihood functions for a given data sample can be calculated
from the following expression
\begin{equation}
    {\cal{L}}_{0,1} = P(N | N_{0,1}) \cdot \prod_{i} f_{0,1}(x^i), 
\end{equation}
where $N$ is the number of events in the data sample, $N_{0,1}$ is the mean number of events for the respective hypothesis, and
$P(N | N_{0,1})$ is the Poisson function. 
The product is taken over all events in the data sample. The function $f_{0,1}(x)$ is the probability density function of the 
studied variables $x = (\theta, \theta_1, \theta_2)$. 

It is seen from Eq.~(12) that the likelihoods are sensitive to the two factors. 
First, the event number in the data sample depends on HNL contribution,
and it is taken into account by the Poisson term. 
Second, the interference between the
HNL and SM diagrams changes the distribution shape, and that is accounted for by the product term in the likelihood function.

To determine the probability density functions, two training data samples are generated. $f_0(x)$ are determined from
the data sample produced without the HeavyN model included in the generator.
The second data sample, used to evaluate $f_1(x)$ is generated with the HeavyN model
included. The mixing parameter and the HNL mass are chosen in the studied region.
After producing the data samples, we apply the cuts described
above and study the three-dimensional distributions of the $\theta, \theta_1, \theta_2$
angles. No correlations between them are found, so
the three-dimensional probability density
function has been chosen as a product of the one-dimensional probability density functions (PDFs)
\begin{equation}
    f(x^i) = f(\theta^i)\cdot f(\theta^i_1)\cdot f(\theta^i_2).
\end{equation}
One-dimensional PDFs are defined as follows
\begin{equation}
    f(\theta) = a(\cos \theta)^4 + b(\cos \theta)^2 + c,
\end{equation}
\vspace{-0.6cm}
\begin{equation}
    f(\theta_{1,2}) = a_{1,2}(\cos \theta_{1,2})^2 + b_{1,2},
\end{equation}
where $a,a_{1,2}, b,b_{1,2},c$ are the free parameters of the fit.

To improve the fit sensitivity to the HNL contribution, the fit is constrained
  to the regions $0.0 \le \cos \theta \le 0.6$ and $0.0 \le \cos \theta_{1,2} \le 0.7$.
Therefore, we fit $\cos \theta$ distributions in the $[0.0, 0.6]$ range
and the $\cos \theta_{1,2}$ distributions in the $[0.0, 0.7]$ range.
The distributions in these regions are smooth and well described by the functions above.
Finally, we perform a three-dimensional fit of the $\cos\theta, \cos\theta_1, \cos\theta_2$ distributions
using RooFit package~\cite{verkerke2003roofittoolkitdatamodeling}.

One more test data sample is generated to determine the likelihood functions and estimate the significance
of the HNL contribution for the chosen mixing parameters and masses.
We fix a specific $M(N)$ value and perform the fit for different values of the
mixing parameter to obtain the significance corresponding to the 90~$\%$ upper limit.
To determine the upper limit as a function of $M(N)$, the same procedure
is repeated for all tested $M(N)$ values.
The results are shown in Fig.~8.

\begin{figure}[h!]
    \includegraphics[width = 0.5\textwidth]{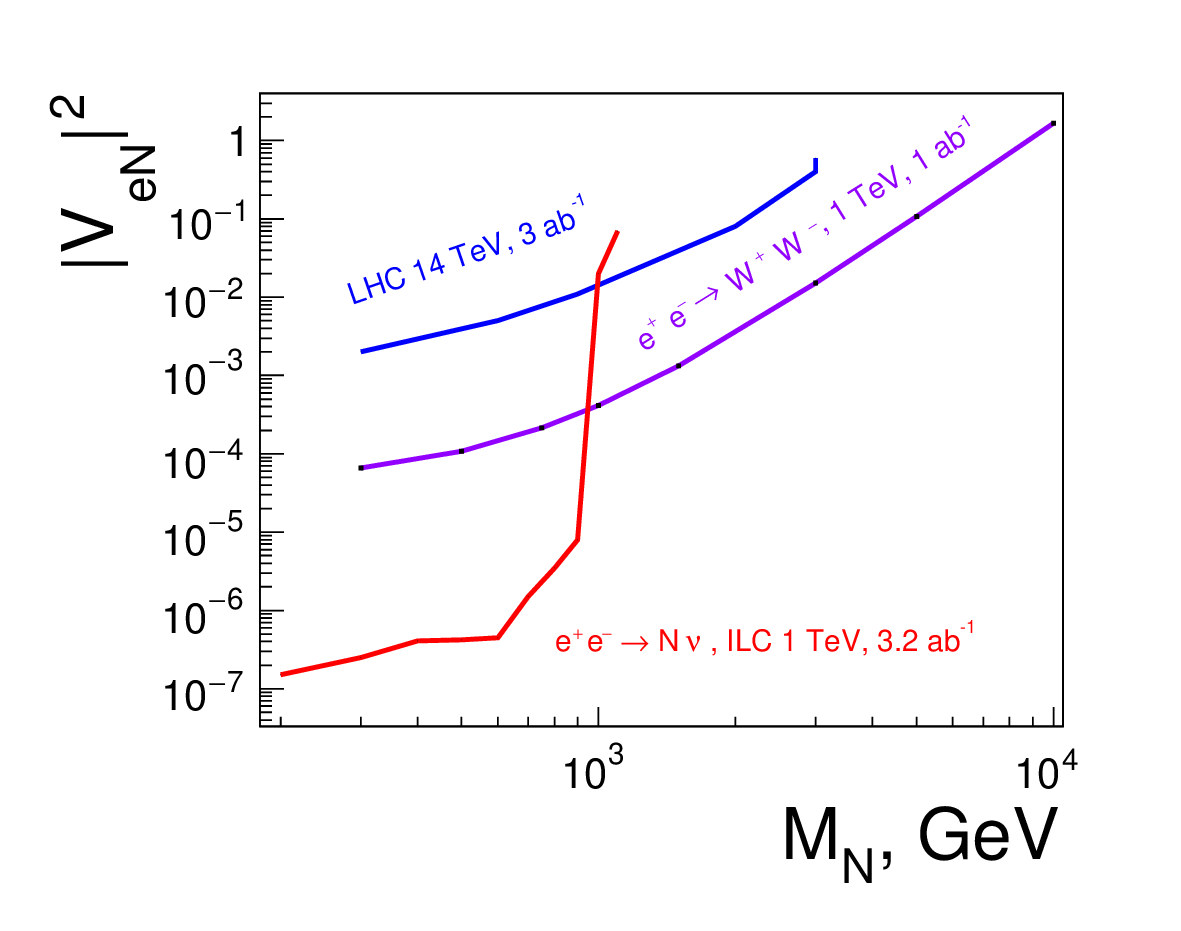}
    \caption{\label{result}The upper limits on the mixing parameter $|V_{eN}|^2$ as a function of $M(N)$. 
    Our results, obtained at $\sqrt{s} = 1~\text{TeV}$ with polarized beams (purple line), are compared
    to the LHC projection for 3 $\text{ab}^{-1}$~\cite{Pascoli_2019} (blue line)
    and to the direct HNL production at 1 TeV~\cite{M_ka_a_2022} (red line).}
\end{figure}

As it is seen in Fig. 8, the obtained upper limits are significantly better than those obtained
at LHC~\cite{Abdullahi_2023} in the region of $300~\text{GeV} < M(N) < 1~\text{TeV}$.
Moreover our results are sensitive to the HNL mass region up to $10^4~\text{GeV}$,
where the upper limit reaches the value $|V_{eN}|^2 \sim 1$.

\section{Conclusions}

We studied the potential heavy neutral lepton contribution to the process $e^+e^- \rightarrow W^+W^-$ within the framework of the seesaw type-I model.
The MC simulation and reconstruction of the signal and background processes is
performed for future $e^+e^-$ colliders at $1~\text{TeV}$. 
The upper limits on the mixing parameter $|V_{eN}|^2$ are obtained as a function of HNL mass $M(N)$.
An experimental study of the $e^+e^- \to WW$ process can significantly
improve the current upper limits in the region of HNL masses larger than $\sqrt{s}$ and effectively
test specific seesaw type-I models. However, the search for direct HNL production can give better upper limits in the region $M(N) < \sqrt{s}$~\cite{M_ka_a_2022}.

It has to be mentioned that there are excellent prospects for the HNL searches in the $e^+e^- \rightarrow W^+W^-$  or $\mu^+\mu^- \rightarrow W^+W^-$ processes at very high collision energies.
With increasing collision energies, the cross section of the SM process $e^+e^- \rightarrow W^+W^-$ quickly decreases, 
whereas the HNL-induced cross section increases. Therefore,
future lepton colliders with the $\sqrt{s} = 3$ and $\sqrt{s} = 10~\text{TeV}$ center of mass energies
can provide strict upper limits on the mixing parameters $|V_{lN}|^2$.
We plan to conduct respective calculations in the future.

\section*{Acknowledgement}

The authors are grateful to Pavel Murat for editing the text.

\end{document}